\documentclass[review]{elsarticle}

\usepackage{lineno,hyperref}
\usepackage{bm}
\usepackage{amsmath}
\usepackage{amssymb}
\usepackage{graphicx}
\usepackage{epstopdf}
\usepackage{upgreek}
\usepackage{multirow}
\usepackage[ruled,lined,commentsnumbered,boxed]{algorithm2e}
\usepackage{epsfig}
\usepackage{psfrag}
\usepackage{subfigure}
\usepackage{array}
\usepackage{threeparttable}
\usepackage{booktabs}
\usepackage{multirow}
\usepackage{longtable}
\modulolinenumbers[5]

\journal{Neural Networks}

\begin{document}
	\renewcommand{\floatpagefraction}{.8}%
	
	\begin{frontmatter}
		
		\title{A Transformer-based deep neural network model for SSVEP classification}
		
		
		\author[aaddress]{Jianbo Chen}
		\author[aaddress]{Yangsong Zhang \corref{corauthor}}
		\author[aaddress]{Yudong Pan}
		\author[baddress]{Peng Xu \corref{corauthor}}
		\author[caddress]{Cuntai Guan}
		
		\address[aaddress]{Laboratory for Brain Science and Medical Artificial Intelligence, School of Computer Science and Technology, Southwest University of Science and Technology, Mianyang, China}
		\address[baddress]{MOE Key Laboratory for NeuroInformation, Clinical Hospital of Chengdu Brain Science Institute, and Center for Information in BioMedicine, School of Life Science and
			Technology, University of Electronic Science and Technology of China, Chengdu, China}
		\address[caddress]{School of Computer Science and Engineering, Nanyang Technological University, Singapore}
		\cortext[corauthor]{Corresponding authors: Yangsong Zhang(zhangysacademy@gmail.com); Peng Xu(xupeng@uestc.edu.cn)}

	\begin{abstract}
		Steady-state visual evoked potential (SSVEP) is one of the most commonly used control signal in the brain-computer interface (BCI) systems. However, the conventional spatial filtering  methods for SSVEP classification highly depend on the subject-specific calibration data. The need for the methods that can alleviate the demand for the calibration data become urgent. In recent years, developing the methods that can work in inter-subject classification scenario has become a promising new direction. As the popular deep learning model nowadays, Transformer has excellent performance and has been used in EEG signal classification tasks. Therefore, in this study, we propose a deep learning model for SSVEP classification based on Transformer structure in inter-subject classification scenario, termed as SSVEPformer, which is the first application of the transformer to the classification of SSVEP. Inspired by previous studies, the model adopts the frequency spectrum of SSVEP data as input, and explores the spectral and spatial domain information for classification. Furthermore, to fully utilize the harmonic information, an extended SSVEPformer based on the filter bank technology (FB-SSVEPformer) is proposed to further improve the classification performance. Experiments were conducted using two open datasets (Dataset 1: 10 subjects, 12-class task; Dataset 2: 35 subjects, 40-class task) in the inter-subject classification scenario. The experimental results show that the proposed models could achieve better results in terms of classification accuracy and information transfer rate, compared with other baseline methods. The proposed model validates the feasibility of deep learning models based on Transformer structure for SSVEP classification task, and could serve as a potential model to alleviate the calibration procedure in the practical application of SSVEP-based BCI systems.
	\end{abstract}


\end{frontmatter}

	\section{Introduction}
	
	Brain-computer interface (BCI) has become a popular research direction in human-computer interaction and medical rehabilitation, which can directly connect the brain to external devices without going through the peripheral nervous system, enabling bidirectional information transmission and feedback \cite{wolpaw2007brain, yao2020bacomics}. Electroencephalogram (EEG)-based BCIs obtain the intentions of the brain through EEG signals, and have attracted attention due to the advantages of convenience, low cost, and non-invasiveness \cite{abiri2019comprehensive}. Among the various EEG paradigms, the high signal-to-noise ratio and low training time of steady-state visual evoked potential (SSVEP) make it one of the most popular paradigms.
	
	SSVEP refers to the EEG in the visual cortex when the subject gazes at a flickering visual stimulus modulated by a constant frequency~\cite{zhang2012Multiple}. The frequencies of SSVEP are the same as the coding frequency of received visual stimuli as well as its harmnoics\cite{regan1989evoked}. By virtue of this characteristic of SSVEP, it is possible to design SSVEP-based BCI system, such as SSVEP-based speller \cite{nakanishi2015comparison}, in which different targets are encoded by different stimulus frequencies. When the subjects need to select a command, they can gaze at the corresponding flickering target stimulus that coding the command on the interface. The generated SSVEP can be identified by a specially designed decoder to obtain the intention of the subject.
	
	In the SSVEP-based BCI system, the robust classification of the SSVEPs is very important \cite{nakanishi2014high}. As the SSVEP frequency is the same as the stimulus frequency, some researches developed the algorithms based on the prior frequency information, such as power spectral density analysis (PSDA) \cite{wang2006practical} and canonical correlation analysis (CCA) \cite{lin2006frequency}, etc. In addition to the fundamental frequency component, SSVEP also contains harmonic components whose frequencies are multiples of the fundamental frequency~\cite{muller2005steady}. Based on this characteristic, filter bank technology was introduced to extend the original CCA (FBCCA)~\cite{chen2015filter}.  FBCCA uses CCA in multiple subbands of SSVEP data, and finally weights the correlation coefficients calculated from these subbands. The FBCCA improves the classification performance by distinguishing the fundamental frequency and harmonics, demonstrating the effectiveness of the filter bank technique on SSVEP classification. Nowadays, filter bank technology has been widely used in various methods~\cite{zhang2019hff,qin2021fbmsi}.
	
	\begin{figure}
		\centering
		\includegraphics[width=4.7in]{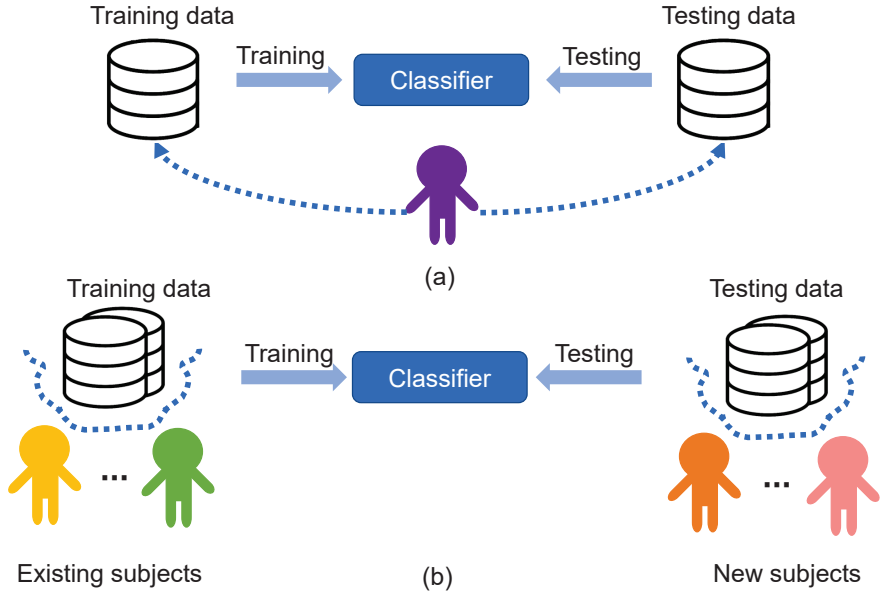} 
		\caption{The diagram of two classification scenarios. (a) intra-subject classification; (b) inter-subject classification.}
		\label{fig:figure0}
	\end{figure}
	
	However, due to the complexity of EEG, SSVEP data always contains noises, such as spontaneous EEG and electromagnetic interference, seriously polluting the signal \cite{hsu2015evaluate}. Traditional training-free methods (such as PSDA, CCA) have better results only when the data length is long. To address the noise interference in SSVEP, a series of recognition algorithms based on machine learning have been proposed. Such methods perform under the intra-subject classification condition, in which the training and testing data are from the same subjects, as shown in Fig.~\ref{fig:figure0}(a). In this condition, the model can obtain parameters that are more suitable for a specific subject, thereby reducing noise interference~\cite{zerafa2018train}. For example, individual template based CCA (IT-CCA) calculates the average of the subject's existing SSVEP signals at each stimulation frequency and uses it as the reference signal for CCA \cite{bin2011high}. This method can add subject-specific patterns to the reference signal, and is widely used in subsequent algorithms. Task-related component analysis (TRCA) method obtains spatial filters by maximizing the reconstitution between SSVEPs of different trials to reduce the noise of SSVEPs and reference signals \cite{nakanishi2017enhancing}. Correlated component analysis (CORCA) learns spatial filters by maximizing the correlation between data to reduce background noise \cite{zhang2018correlated}. Task-discriminant component analysis (TDCA) uses multi-class linear discriminant analysis to learn spatiotemporal filters and classify in a discriminant manner \cite{liu2021improving}.
	
	The above method has significant effect in the intra-subject classification experiment, in which the training data and the testing data belong to the same subject \cite{ravi2020comparing}. However, the collection of SSVEP data is a time-consuming and laborious work. Hence, a potential and challenging direction is to transfer the data from existing subjects to new subjects in the inter-subject classification scenario, under which a classifier can be obtained with the data from already existing subjects and then used the classifier to test the data from new subjects, as shown in Fig.~\ref{fig:figure0}(b). Although many works have improved traditional state-of-the-art methods to adapt them to inter-subject scenario, the results may be not optimal~\cite{yan2022cross}. Because the brain processes natural sensory stimuli in a dynamic, non-fixed, and nonlinear manner, SSVEP is non-stationary and varies widely among individuals~\cite{ibanez2019characterization}. Even the data collected by the same subject, the data acquired at different times may also have different distribution. These situations pose great challenges for inter-subject experiments, and the performance of traditional machine learning-based algorithms under inter-subject condition degrades greatly, which is far from its performance under intra-subject condition.
	
	In recent years, deep learning has been developed significantly and has made milestone progress in areas such as computer vision and natural language processing \cite{krizhevsky2017imagenet, collobert2011natural}. Deep learning models have powerful feature extraction capabilities and can directly be applied on the raw data\cite{craik2019deep, schmidhuber2015deep}. Deep learning models have been used on many EEG classification tasks, including convolutional neural networks (CNN) \cite{zhou2018epileptic}, recurrent neural networks (RNN) \cite{du2020efficient}, graph neural networks (GNN) \cite{zhong2020eeg}, etc. Several studies have used deep learning to process SSVEP data, achieving outstanding performance on classification tasks, especially inter-subject classification. For instances, EEGNet is a compact convolutional neural network (CNN) that uses CNNs to implement the spatial-temporal filtering and feature extraction, achieving significantly better results than traditional methods under inter-subject conditions \cite{waytowich2018compact}. The idea of using temporal and spatial convolutions has achieved promising results, which has also influenced many later algorithms. The subject invariant SSVEP generative adversarial network (SIS-GAN) uses generative adversarial networks to generate artificial SSVEP data  to expand the training dataset  \cite{aznan2021leveraging}. Complex convolutional neural network (CCNN) uses the complex spectrum of SSVEP signal as the input of CNN for classification, demonstrating the effectiveness of complex spectral features on SSVEP classification \cite{ravi2020comparing}. InceptionEEG-Net (IENet) combines Inception with residual connections and uses multi-scale convolution kernels to extract features from receptive fields of different sizes \cite{du2022ienet}. In addition, filter bank technology is also applied in deep learning models to extend the existing models, such as FB-EEGNet and FBCNN~\cite{yao2022fb, zhao2021filter}.
	
	Although the deep learning-based SSVEP recognition model has made great progress compared with the traditional machine learning algorithm in the inter-subject environment, it still has large improvement space to meet the actual needs for SSVEP-based BCI. Under the inter-subject condition, the model should have good generalization performance for unseen subjects. In addition, due to the black-box nature of deep learning, the interpretability of existing deep learning-based models still need to investigate for explaining the classification mechanisms as the traditional methods. In recent years, Transformer becomes one of the most promising model structures, which was first used in machine translation and quickly took natural language processing by storm with its excellent performance \cite{devlin2018bert,vaswani2017attention}. It was then applied to the field of computer vision and achieved brilliant results \cite{dosovitskiy2020image}. So far, Transformer based models become very powerful in many fields with wide applicability, and are more interpretable compared with other neural networks\cite{vig2019multiscale}. Transformer has excellent feature extraction ability, and the extracted features have better performance on downstream tasks. In BCI systems, some studies have applied Transformers to the EEG classification tasks, and achieved good results \cite{xie2022transformer, guo2022transformer, phan2022sleeptransformer}. To the best of our knowledge, the Transformer has not been leveraged for SSVEP classification.
	
	In this paper, a deep learning model for SSVEP classification based on Transformer structure, termed as SSVEPformer, is proposed.  Inspired by previous study~\cite{ravi2020comparing}, the complex spectra of the SSVEP data were adopted as the input of the  SSVEPformer model, allowing the model to focus on the frequency domain property of SSVEP data. In addition, we presented a extended SSVEPformer based on the filter bank technology, termed as FB-SSVEPformer, to further improve the classification performance by fully utilizing the harmonic information. To validate the performance of the proposed and compared models, we utilized two public SSVEP datasets. Dataset 1 has 12 categories from ten subject~\cite{nakanishi2015comparison}, and Dataset 2 has 40 categories from 35 subjects~\cite{wang2016benchmark}. Using 1 s time window, FB-SSVEPformer achieves 88.37$\%$ accuracy and 112.45 bits/min information transfer rate (ITR) on Dataset 1, and 83.19$\%$ accuracy and 157.65 bits/min ITR on Dataset 2, under inter-subject classification scenario.
	
	The rest of the paper is structured as follows: Section 2 introduces the datasets and baseline methods, and presents the proposed SSVEPformer and FB-SSVEPformer. The performance of the baseline methods and the proposed method on the two datasets is presented in the Section 3. Section 4 discusses the performance and limitations of the model. Finally, Section 5 presents the conclusion for this study.
	
	\section{Materials and methods}
	
	\subsection{Datasets}
	
	Two public datasets were adopted to evaluate the performance of all the methods.
	
	Dataset 1 \cite{nakanishi2015comparison}: This dataset was acquired with 12 visual target stimuli on a 27-inch
	LCD monitor, which were modulated by the frequencies ranged from 9.25 Hz to 14.75 Hz with 0.5 Hz steps. Ten subjects with normal or corrected-to-normal vision participated in the experiment, sitting in chairs 60 cm from the monitor in a dimly lit room. The BioSemi ActiveTwo EEG system (Biosemi, Inc.) was used to acquire EEG data from eight electrodes in the occipital region.The whole experiment consisted of 15 blocks. In each block, there were 12 trials corresponding to the 12 flickering target stimuli, in each of which the subjects were required to gaze at one of the 12 target stimuli. The gazed target stimulus in each trial was random selected, and each trial lasted for 4 seconds. The EEG signal was sampled at a sampling rate of 2048 Hz, and then downsampled to 256 Hz.
	
	Dataset 2~\cite{wang2016benchmark}: The dataset consisted of 40 visual target stimuli displayed on an LCD monitor. The 40 targets were encoded using the joint frequency and phase modulation (JFPM) method, with target stimulation frequencies ranged from 8 Hz to 15.8 Hz with 0.2 Hz step, and phases starting at 0 with 0.5 $\pi$ steps. Thirty-five subjects with normal or corrected-to-normal vision participated in the experiment, eight of whom had experience using SSVEP-based spell. For each subject, the experiment consisted of 6 blocks. In each block, the subjects stared at 40 targets in random order according to the prompts, resulting in a total of 40 trials. Each trial lasted for 6 seconds. For the first 0.5 seconds, the subjects were asked to move the realization to the target stimulus position according to the prompt, then fixated on the target stimulus for 5 seconds, and finally the monitor was blank for 0.5 seconds. EEG data were acquired using a Synamps2 EEG system (Neuroscan, Inc.) at a sampling rate of 1000 Hz and down-sampled to 250 Hz. Finally, a 50 Hz notch filter was used to eliminate power frequency interference.
	
	\subsection{Data preprocessing}
	
	As in the previous study, all eight channels (O1, Oz, O2, PO3, POZ, PO4, PO7, PO8)  in Dataset 1 and  nine channels (O1, Oz, O2, PO3, POZ, PO4, PZ, PO5 and PO6) in Dataset 2 were adopted for classification~\cite{ravi2020comparing,nakanishi2017enhancing}. For the models that need time-domain data as input, fourth-order Butterworth filter with 8-64 Hz bandpass range are used to filter the data and take the filtered data as input. Suppose the time window length for classification be $d~s$, considering the visual delay, the data segments in Dataset 1 were extract in the time window $[0.135~s, d+0.135~s]$  after the stimulus onset~\cite{nakanishi2015comparison}; the data segments were extracted in the time window $[0.64~s, d+0.64~s]$ in the Dataset 2~\cite{wang2016benchmark}.
	
	For models that require frequency domain data as input, FFT was used to convert EEG data in the time domain into the frequency domain. The result of the FFT can be expressed as:
	
	\begin{equation}
		FFT(x)=Re[FFT(x)]+iIm[FFT(x)]
	\end{equation}
	where $x$ represents the preprocessed EEG data in the time domain, $i$ is the imaginary unit, $Re$ and $Im$ represent the real and imaginary parts of complex spectrum, respectively. For frequency domain data, there are two ways to transform it into model input, namely magnitude spectrum $ X_{mag} $ and complex spectrum $ X_{comp} $~\cite{ravi2020comparing}:
	
	\begin{flalign}
		& X_{mag}=\sqrt{\{Re[FFT(x)]\}^{2}+\{Im[FFT(x)]\}^{2}} \\
		& X_{comp}=Re[FFT(x)]||Im[FFT(x)]
	\end{flalign}
	
	The symbol $||$ denotes the concatenation operation. The magnitude spectrum calculates the sum of the squares of the real and imaginary parts at each frequency point, discarding the phase information of the data and only containing the magnitude information. The complex spectrum concatenates the real and imaginary parts of the complex Fourier spectrum, and both magnitude and phase information are preserved. Previous studies have shown that phase information has a role in SSVEP classification \cite{pan2011enhancing, chen2014hybrid}, and complex spectrum input also outperforms magnitude spectrum in comparative experiments~\cite{ravi2020comparing}. Therefore, in this study, we used the complex spectrum $ X_{comp} $ of the data as the input of the proposed model. The complex spectra input denoted as $ I_{comp} $, is defined as:
	
	\begin{equation}
		I_{comp}=\begin{bmatrix}X_{comp}(CH_{1})
			\\X_{comp}(CH_{2})
			\\\vdots
			\\X_{comp}(CH_{n})
			
		\end{bmatrix}
	\label{equ:equation4}
	\end{equation}
	where $CH_1, CH_2, \cdots, CH_{n}$ represent different EEG channels. Specifically, for dataset 1, the time-domain input data is padded with zeros so that the resolution after FFT is 0.25 Hz; for dataset 2, zero padding is still used so that the frequency resolution after FFT is 0.2 Hz. After that, we selected the complex spectrum in the range of 8 to 64 Hz, and concatenated the real part and the imaginary part as the input of SSVEPformer and FB-SSVEPformer.
	
	\subsection{The baseline methods}
	\subsubsection{TRCA}
	
	TRCA is a spatial filtering method that learns the spatial filters by maximizing the reproducibility between existing data of the same class, which can extract task-related components in the data~\cite{nakanishi2017enhancing}. The average of existing data of the same class is then used as the reference signal for that class. With the spatial filters obtained by TRCA based on the calibration data, the correlation coefficients between the projected features of the test sample and various reference signals can be computed, and then obtained the classification result. In the experiment of this study, owing to the inter-subject classification scenario, the data of one test subject will be selected as the test sampels, and the data of all other subjects were pooled together as the calibration data to calculate the spatial filter and the reference signals of each stimulus frequency.
	
	\subsubsection{EEGNet}
	
	EEGNet is a deep learning model specially designed for EEG signal processing, and has been widely used in various EEG classification tasks since it was proposed, such as motor imagery, P300, SSVEP, etc \cite{waytowich2018compact,lawhern2018eegnet}. Waytowich et al. applied EEGNet to SSVEP classification and achieved excellent results in the inter-subject classification task~\cite{waytowich2018compact}. EEGNet consists of four layers. The first layer is a convolutional layer that simulates filtering operation on the EEG data in each channel. The second layer is a depth-wise convolutional layer, which is equivalent to a spatial filter that weights all the channels. The third layer is a separable convolutional layer for extracting categorical features. The fourth layer is a fully connected layer, which outputs the classification result.
	
	\subsubsection{CCNN}
	
	CCNN transforms the SSVEP to the frequency domain using FFT, using the complex spectrum of the signal as the input of the model in order to extract frequency and phase information~\cite{ravi2020comparing}. The model consists of convolutional layers and fully connected layers. Convolutional layers are responsible for spatial filtering, temporal filtering, and feature extraction. The fully connected layer summarizes the extracted features to get the final result. CCNN achieves excellent results using complex spectrum input, proving that complex spectrum representation can be beneficial for SSVEP classification.
	
	\subsection{The proposed SSVEPformer}
	
	\begin{figure*}
		\centering
		\includegraphics[width=4.7in]{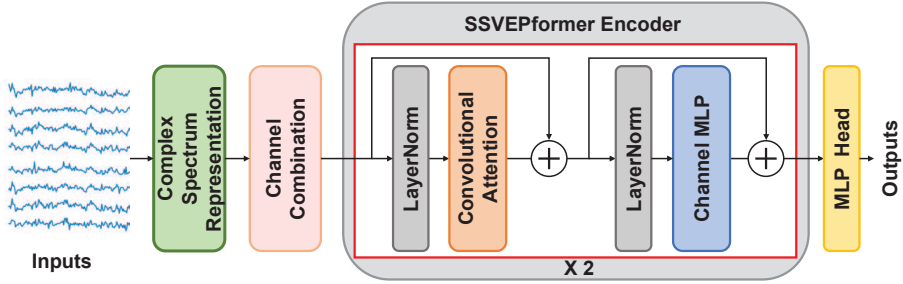} 
		\caption{The architecture of SSVEPformer model.  SSVEPformer consists of six blocks: the input, complex spectrum representation, channel combination, SSVEPformer encoder, MLP head and the output. The ``$\times 2$'' means that two identical and successive operation sub-encoders shown in the red rectangle box constitute the SSVEPformer encoder.} 
	    \label{fig:figure1}
	\end{figure*}
	
	The model adapts to the unique characteristics of SSVEP as much as possible on the basis of the Transformer structure, and can be considered as a Transformer-esque SSVEP recognition model. Fig.~\ref{fig:figure1} illustrates the model architecture diagram. The SSVEPformer consists of six blocks: the input, the complex spectrum representation, channel combination, SSVEPformer encoder, multilayer perceptron (MLP) head and the output. In the complex spectrum representation block, the input EEG was transformed into complex spectrum $ I_{comp} $ as defined in formula~(\ref{equ:equation4}). As the core components of SSVEPformer, the detailed description of channel combination, SSVEPformer encoder and MLP head is as follows, and the detailed structure is shown in Table 1. In the following description, $N$ is denoted as the total number of classes, $C$ is the number of SSVEP channels, and $F$ is the length of the complex spectrum in each channel.

	\begin{table*}
		\centering
		\caption{Detailed architecture and parameters in the core components of SSVEPformer.}
		\resizebox{\textwidth}{!}{%
			\begin{tabular}{c|c|c|c|c|c}
				\hline
				Block     &\multicolumn{2}{c|}{Module}     &Layer     & Output size       & Explanation                    \\ \hline
				\multirow{5}{*}{Channel combination}    & \multicolumn{2}{c|}{\multirow{5}{*}{}}                                     & Input      & (C, F)      &                                                    \\
				& \multicolumn{2}{c|}{}                                                      & Conv1d     & (2 $\times$ C, F)  & filters = 2 $\times$ C, kernalsize = 1, padding = 'same'  \\
				& \multicolumn{2}{c|}{}                                                      & LayerNorm  & (2 $\times$ C, F)  &                                                    \\
				& \multicolumn{2}{c|}{}                                                      & Activation & (2 $\times$ C, F)  & GELU                                               \\
				& \multicolumn{2}{c|}{}                                                      & Dropout    & (2 $\times$ C, F)  & dropoutrate = 0.5                                  \\ \hline
				\multirow{11}{*}{SSVEPformer encoder} & \multirow{11}{*}{Sub-encoders ($\times$ 2)} & \multirow{6}{*}{CNN module}           & LayerNorm  & (2 $\times$ C, F)  &                                                    \\
				&                                    &                                      & Conv1d     & (2 $\times$ C, F)  & filters = 2 $\times$ C, kernalsize = 31, padding = 'same' \\
				&                                    &                                      & LayerNorm  & (2 $\times$ C, F)  &                                                    \\
				&                                    &                                      & Activation & (2 $\times$ C, F)  & GELU                                               \\
				&                                    &                                      & Dropout    & (2 $\times$ C, F)  & dropoutrate = 0.5                                  \\
				&                                    &                                      & Residual1  & (2 $\times$ C, F)  & adding the input of this module                    \\ \cline{3-6}
				&                                    & \multirow{5}{*}{Channel MLP module} & LayerNorm  & (2 $\times$ C, F)  &                                                      \\
				&                                    &                                      & Linear     & (2 $\times$ C, F)  &                                          \\
				&                                    &                                      & Activation & (2 $\times$ C, F)  & GELU                                               \\
				&                                    &                                      & Dropout    & (2 $\times$ C, F)  & dropoutrate = 0.5                                  \\
				&                                    &                                      & Residual2  & (2 $\times$ C, F)  & adding the input of this module                     \\ \hline
				\multirow{7}{*}{MLP head}             & \multicolumn{2}{c|}{\multirow{7}{*}{}}                                     & Flatten    & (2 $\times$ C $\times$ F) &                                                    \\
				& \multicolumn{2}{c|}{}                                                      & Dropout    & (2 $\times$ C $\times$ F) & dropoutrate = 0.5                                  \\
				& \multicolumn{2}{c|}{}                                                      & Linear     & (6 $\times$ N)     &                                      \\
				& \multicolumn{2}{c|}{}                                                      & LayerNorm  & (6 $\times$ N)     &                                                    \\
				& \multicolumn{2}{c|}{}                                                      & Activation & (6 $\times$ N)     & GELU                                               \\
				& \multicolumn{2}{c|}{}                                                      & Dropout    & (6 $\times$ N)     & dropoutrate = 0.5                                  \\
				& \multicolumn{2}{c|}{}                                                      & Linear     & (N)         &                                       \\ \hline
			\end{tabular}%
		}
	\end{table*}

	\subsubsection{Channel combination block}
	
	The SSVEP data  were usually acquired from multiple channels, and these data not only contain valuable SSVEP classification information, but also include various artifacts that interfere with the classification. Since the channels are distributed at different locations on the scalp of brain, the background components may be different. Using the channel combination to calculate the weighted combination of all the channels can suppress the noise and enhance the SSVEP component that helps the classification. In addition, the channel combination will be used multiple times to obtain multiple channel weighted combinations. This is because different combination methods focus on different classification information or suppress different noises, and multiple combination operations can improve the robustness and performance of the model. The channel combination block uses convolutional layers to perform weighted combination between channels. The convolutional layer convolves the number of data channels from $C$ to $2 \times C$ using a convolution kernel of length 1 with Conv1d function in Pytorch framework. This process is equivalent to learning $2 \times C$ spatial filters to weight the data of each channel.
	
	\subsubsection{SSVEPformer encoder block}
	The original Transformer encoder generally consists of two components, an attention module and a channel MLP module~\cite{vaswani2017attention}. The former is used to mix information between tokens, and the latter includes channel MLP for feature extraction. Although the attention mechanism was widely used in the original Transformer, many subsequent studies have found that the attention mechanism is not necessary for the Transformer \cite{tay2021synthesizer}. Some studies have used convolution, MLP or even pooling to replace the attention module and achieved similar results \cite{yu2022metaformer}. Similarly, SSVEPformer keep the channel MLP module unchanged and replace the attention module with a CNN module.
	
	The SSVEPformer encoder consists of two identical and successive sub-encoders as shown in Fig.~\ref{fig:figure1}. Each sub-encoder consists of two modules, the first is the CNN module and the second is the channel MLP module. Residual connection is used in each module, the input of each module is also added to the output of this module. In the CNN module, the Conv1d layer extracts features in the channel dimension using a convolution kernel of length 31, and fuses the features of each channel. In the channel MLP module, features are extracted on each channel using a linear transformation. This linear transformation applies the same operation to each channel, mapping from $F$ elements to new $F$ elements, using high computational cost to obtain global fine-grained features.
	
	\subsubsection{MLP head block}
	
	The features extracted by the upstream blocks are finally input into the MLP head block, which is mainly composed of two fully connected layers. The input data is first flattened to facilitate subsequent operations. Followed by two fully connected layers, the data length is gradually mapped to the number of categories of the dataset. Considering that the input data is already the features extracted by the previous blocks, in order to avoid losing useful information, the MLP head uses two linear layers to continuously refine the input features to obtain the final result.
	
	The LayerNorm layer, GELU layer, and Dropout layer used in the network are all common components of deep neural networks. The LayerNorm layer regularizes the data in the channel dimension to ensure the stability of the data feature distribution, making training faster and more stable \cite{xu2019understanding}. The GELU layer adds nonlinear operations to the network, making the model capable of nonlinear fitting \cite{hendrycks2016gaussian}. Meanwhile, GELU has been proven to be a high-performance activation function and one of the current state-of-the-art activation functions~\cite{zhang2021study}. Dropout can avoid overfitting during model training and improve model generalization \cite{baldi2013understanding}.
	
	
	\subsection{FB-SSVEPformer}
	
	In the SSVEP data, there are harmonic components whose frequencies are multiple times of the fundamental frequency, which can also be used as classification features. For example, FBCCA filters the data into several subbands and actively guides the model to pay attention to the information of harmonics, and achieves better results than original CCA~\cite{chen2015filter}.  Here, we presented a extended SSVEPformer based on the filter bank technology (FB-SSVEPformer) to further improve the classification performance by fully utilizing the harmonic information. The model structure is shown in Fig.~\ref{fig:figure2}. The original EEG data are filtered into $S$ frequency subbands, which are then fed to the corresponding subnetworks as input. FB-SSVEPformer uses SSVEPformer as a subnetwork, and each subnetwork handles data of different subbands. The results of each subnetwork($ \rho_{s,f}, s=1,2,\cdots,S, f=1,2,\cdots,N$) are finally weighted and fused to obtain the final result $\mathop{Y_f}$ as follows:
	
	\begin{figure*} 
		\centering
		\includegraphics[width=4.7in]{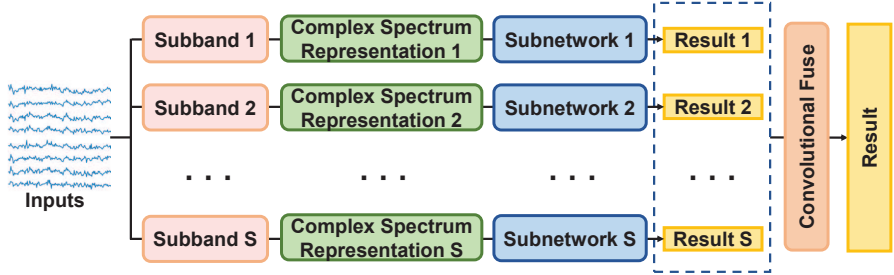} 
		\caption{The architecture of FB-SSVEPformer, where $S$ is the number of subbands and the subnetwork is SSVEPformer. The input data is transformed into the data of $S$ subbands through the filter bank, and then transformed into the frequency domain through FFT. Each subnetwork uses the data of corresponding subband to obtain the result, and then obtains the final result through the convolutional fusion operation.} 
	\label{fig:figure2}
	\end{figure*}
	
	\begin{equation}
		\sideset{_{}^{}}{_{}^{}}{\mathop{Y_f}}=\sideset{_{}^{}}{_{}^{}}{\mathop{argmax}}_{f={1,...,N}}^{}\sum_{s=1}^{S}w_{s} \cdot \rho_{s,f}
	\end{equation}
	where $N$ is the total number of categories, $w_{s}( s=1,2,\cdots,S)$ is the weighted parameters of the convolution kernel, and $S$ is the number of subbands. The fusion operation refers to the method of FBCCA fusing the results of each subband, which gives special weights to the results of each subband, and then weights them as the final result. However, the weights in FBCCA are constants obtained by grid search~\cite{chen2015high}. FB-SSVEPformer uses convolutional layers to implement weighting operation, and the parameters move in the direction of the best classification result during training and eventually stabilize.
	
\begin{figure} 
		\centering
		\includegraphics[width=4.7in]{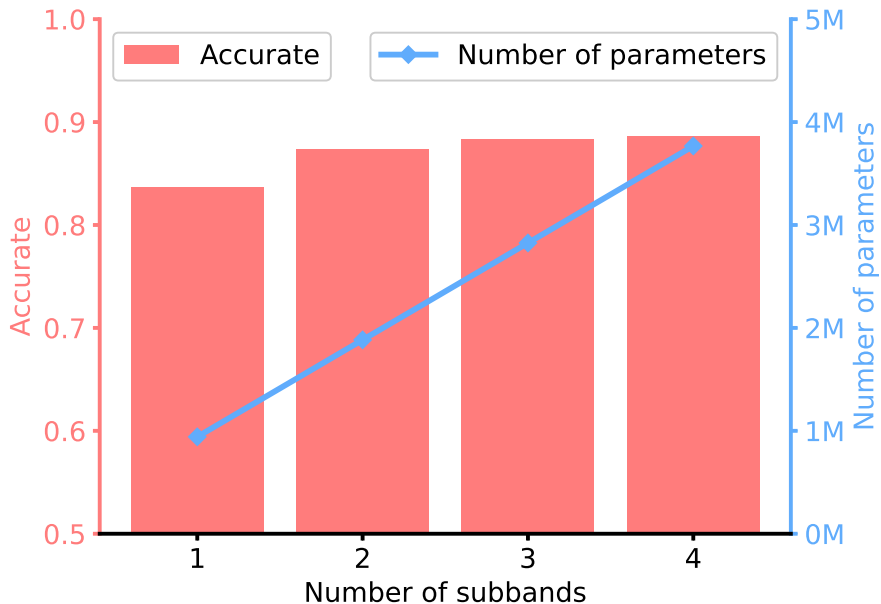} 
		\caption{The relationship between the number of subbands of the FB-SSVEPformer and the accuracy and parameters of the model. The experiment uses Dataset 1 with the data length of 1 s. The bars represent the average precision, the lines represent the number of parameters, and the 'M' means the magnitude of millions.}
	\label{fig:figure3}
	\end{figure}

	For the selection of subbands, the lower and upper cutoff frequencies of the $m-th$ subband were set to $ m \times 8$ Hz and 80 Hz on Dataset 1, and $m \times 9$ Hz and 80 Hz on Dataset 2. The effect of the subband division method has been verified~\cite{chen2015filter,zhang2018correlated,nakanishi2017enhancing}. During the implementation of band-pass filtering, the lower limit of each subband was subtracted by an additional 2 Hz bandwidth~\cite{chen2015filter}. In order to verify the relationship between the number of subbands and the classification effect, FB-SSVEPformer with different numbers of subbands uses data with a data length of 1 s to conduct experiments on Dataset 1, and the results are shown in Fig.~\ref{fig:figure3}. When the number of subbands is 4, the accuracy of the model is slightly improved, but the number of parameters is greatly increased compared to when the number of subbands is 3. When the number of subbands is 2, the accuracy of the model is lower than that when the number of subbands is 3. Considering the balance between the performance and model parameters, the number of subbands in the FB-SSVEPformer model is set to 3 in following analysis.
    
	\subsection{Experimental settings and performance evaluation}
	
	All the deep learning models were implemented with the Pytorch framework. For the SSVEPformer and FB-SSVEPformer models, the convolutional and linear layer parameters were initialized with a normal distribution with mean 0 and variance 0.01. During training, cross-entropy was used to calculate the loss, and backpropagation was used to update the parameters. Both models used a stochastic gradient descent(SGD) algorithm to compute parameter updates with the learning rate of 0.001, the momentum of 0.9, and an L2 regularization penalty of 0.001. The batch size was set to 128, and the dropout rate is set to 0.5. Notably, for SSVEPformer, the number of training epochs is 100. For FB-SSVEPformer, the model first uses data from different frequency bands to train each subnetwork for 100 epochs to stabilize their results, and then trains the entire model for 20 epochs.
	
	Three models of EEGNet, CCNN, TRCA were used as baseline methods to compare with the proposed two models. For the fair comparison, the preprocessing procedures of the input data are the same as the operation described above.  For CCNN method, the EEG data need further transform into complex spectrum~\cite{ravi2020comparing}.
	
	The classification accuracy and ITR were used as metrics to evaluate the performance of each method. The accuracy is the ratio of the number of the correctly classified samples to the number of the total test samples. ITR (bits/minute) is calculated as follows~\cite{wolpaw2002}:
	
	\begin{equation}
		ITR=\frac{60}{T} \times \left[log_{2}N + Plog_{2}P + (1 - P)log_{2}\frac{1 - P}{N - 1} \right]
	\label{equ:itr}
	\end{equation}
	where $N$ is the total number of categories and $T$ is the average time (seconds) for selection. During calculating the ITR, 0.5 s gaze movement time was added into the parameter $T$ as previous studies~\cite{nakanishi2017enhancing, chen2015high}. For example, when the data length is 1 s, then the T is set to 1.5 s during the ITR calculation with the formula~(\ref{equ:itr}).

	In current study, we focused on the inter-subject classification experiment, the leave-one-subject-out(LOSO) strategy was adopted. Specifically, when evaluating on each of the two datasets, the data of one subject was used as the test set, and the data of all other subjects were used for training set. The procedure repeated until all the subjects were served as the test subject once.
	
	\section{Result}
	
	In order to evaluate all the methods, the experiments were conducted on Dataset 1 and Dataset 2. The TRCA, EEGNet and CCNN were used as the baseline methods, which have achieve excellent performance in previous studies~\cite{nakanishi2017enhancing,ravi2020comparing,waytowich2018compact}.
	
	\subsection{Dataset 1}
	
	\begin{figure*}
	\centering
		\includegraphics[width=4.7in]{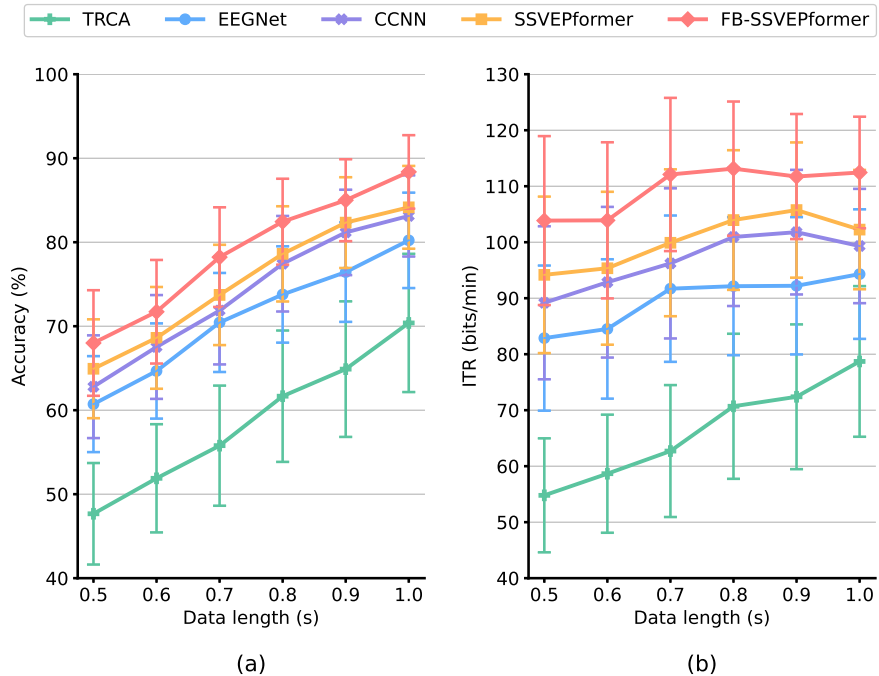} 
		\caption{The average classification results across subjects of the five methods with different data length on Dataset 1. (a) average accuracies; (b) average ITRs. The error bars indicate the standard errors of each results.}
	\label{fig:figure4}
	\end{figure*}
	
	Fig.~\ref{fig:figure4}(a) and (b) show the average classification accuracies and ITRs of the five methods on Dataset 1, respectively. The data length ranges from 0.5 s to 1 s with an interval of 0.1 s. The results show that the proposed SSVEPformer and FB-SSVEPformer outperform other compared baseline methods. FB-SSVEPformer achieves the best results in both average classification accuracies and ITRs, and SSVEPformer achieves better results than the three baseline methods, especially when the data length is shorter than 1 s. From 0.5 s to 1 s, the average accuracies of SSVEPformer are 64.93$\%$, 68.61$\%$, 73.72$\%$, 78.61$\%$, 82.33$\%$ and 84.16$\%$, and the ITRs are 94.17 bits/min, 95.35 bits/min, 99.90 bits/min, 103.96 bits/min, 105.74 bits/min and 102.25 bits/min, respectively. From 0.5 s to 1 s, the average accuracies of FB-SSVEPformer are 68.00$\%$, 71.72$\%$, 78.22$\%$, 82.44$\%$, 85.00$\%$, and 88.37$\%$, and the ITRs are 103.86 bits/min, 103.91 bits/min, 112.10 bits/min, 113.14 bits/min, 111.73 bits/min and 112.45 bits/min, respectively.
	
	\subsection{Dataset 2}
	
	\begin{figure*} 
	\centering
	\includegraphics[width=4.7in]{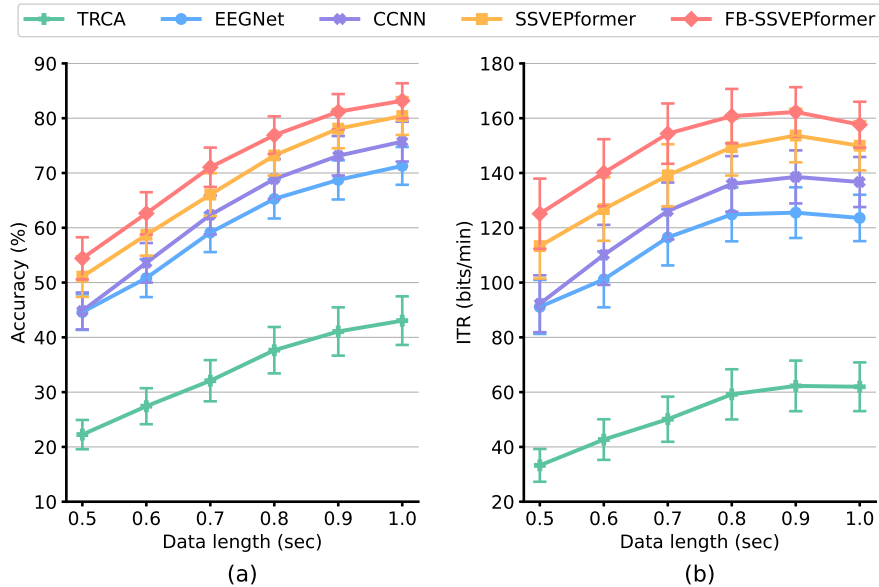} 
		\caption{The average classification results across subjects of the five methods with different data length on Dataset 2. (a) average accuracies; (b) average ITRs. The error bars indicate the standard errors of each results.}
	\label{fig:figure5}
	\end{figure*}
	
	Fig.~\ref{fig:figure5}(a) and (b) show the average classification accuracies and ITRs of the five methods on Dataset 2, respectively. As the results on Dataset 1, the experimental results show that the SSVEPformer and FB-SSVEPformer outperform other three methods. FB-SSVEPformer achieves the best results in both average classification accuracies and ITRs, and SSVEPformer is also better than the baseline methods. From 0.5 s to 1 s, the average accuracies of SSVEPformer are 51.08$\%$, 58.71$\%$, 66.06$\%$, 73.20$\%$, 78.10 $\%$, and 80.40$\%$, and the ITRs are 113.31 bits/min, 126.83 bits/min, 139.14 bits/min, 149.41 bits/min, 153.68 bits/min and 149.95 bits/min, respectively. From 0.5 s to 1 s, the average accuracies of FB-SSVEPformer are 54.42$\%$, 62.63$\%$, 71.05$\%$, 76.90$\%$, 81.18$\%$ and 83.19$\%$ and the ITRs are 125.13 bits/min, 140.10 bits/min, 154.36 bits/min, 160.76 bits/min, 162.28 bits/min and 157.65 bits/min, respectively.
	
	Based on the results on the two widely used datasets for the method evaluation, we could find that the SSVEPformer and FB-SSVEPformer promote the SSVEP classification performance in the inter-subject scenario.
	
	\section{Discussion}
	
	\subsection{A feasible method in inter-subject scenario for SSVEP classification}
	
	The SSVEP-based BCI could provide enough number of targets to code the commands or characters for the application. Although lots of methods have been proposed~\cite{zerafa2018}, it is still challenging to recognize the target with the EEG signals for the users, especially when dealing with the large number of targets. In recent years, the spatial filtering methods based on the calibration data become a popular solution to achieve high classification accuracy, such as the TRCA method and its variant. Whereas, the collection of calibration data is time-consuming and laborious. Therefore, a potential strategy is to utilize the data from existing subjects to train the method for the new subjects in a inter-subject classification scenario that can realize the plug-and-play application without new data collection and calibration procedure. Unfortunately, the data distribution may vary largely among different subjects, which result in the performance of traditional algorithms under inter-subject scenario degrades greatly. These situations pose great challenges for inter-subject classification with the traditional method, such as TRCA~\cite{ravi2020comparing}.
	
	In recent years, deep learning methods have achieved great success for brain signals analysis~\cite{zhang2021}. Deep learning-based solutions for alleviating the calibration data under different BCI paradigms have gradually increased in recent years. For the SSVEP paradigm, the CCNN model provide a option for inter-subject classification without the calibration data from a new subject~\cite{ravi2020comparing}. Even so, it still has large improvement space to meet the practical needs. In recent years, the Transformer and the variants have achieved state-of-the-art performance in various fields~\cite{Han2022pami, lin2021survey}. In this study, we attempted to design a deep learning model with Transformer structure for SSVEP classification in calibration-free data condition. We proposed SSVEPformer and a variant FB-SSVEPformer with filter bank technology, which is the first application of the Transformer to the SSVEP classification. Evaluated on two public datasets with different numbers of targets, with data length of 1 s, the SSVEPformer obtains 84.16 $\%$ and  80.40 $\%$ accuracies, and the FB-SSVEPformer obtains 88.37 $\%$ and 83.19 $\%$ accuracies, on 12-class and 40-class classification task, respectively. The experimental results show that the proposed models could achieve excellent performance in inter-subject classification task. The proposed model validates the feasibility of deep learning models based on Transformer structure for calibration-free SSVEP classification task, and could serve as a potential model to alleviate the calibration procedure in the practical application of SSVEP-based BCI systems.
	
	\subsection{Enhancing the performance with subject-adaptive strategy}
	
	\begin{figure*}
	\centering 
	\includegraphics[width=4.7in]{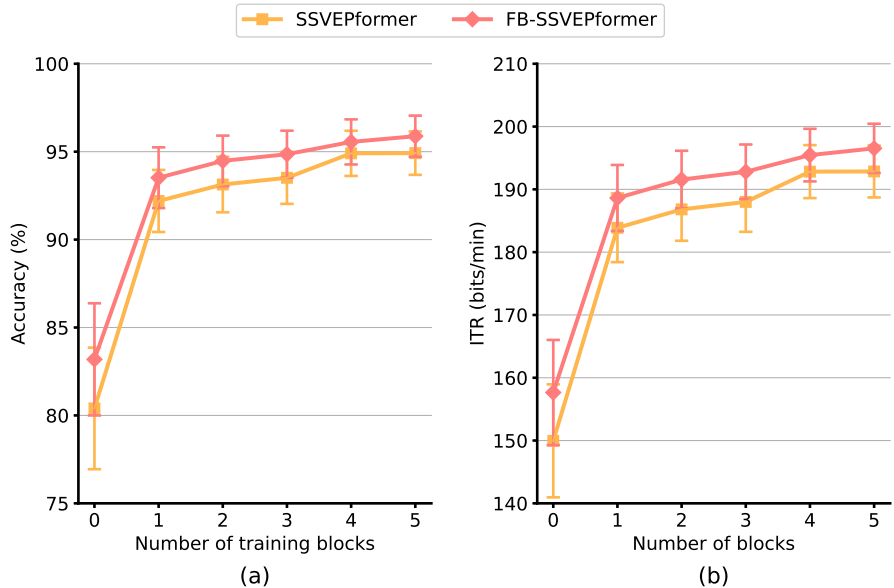} 
		\caption{Subject-adaptive experimental results of SSVEPformer and FB-SSVEPformer. The experiments were coducted on Dataset 2 with the duration of 1 second. The horizontal axis is the amount of data used for subject-adaptive training. When the number of blocks is 0, it means that no subject-adaptive training was performed.}
	\label{fig:figure6}
	\end{figure*}
	
	In recent years, deep learning-based methods have been widely used in SSVEP classification, showing no less performance than traditional methods in both classification performance and generalization. In the SSVEP-based BCI system, when the user uses it for a period of time, some of the user's own data could be collected. These data can be used to update the model for adapting to the user's data and improve the performance of the system. This method of adding part of the data from the test subject in inter-subject classification scenario is subject-adaptive strategy~\cite{zhangmi20211}. Here, in order to investigate the performance of the proposed method under the subject-adaptive strategy, we use the data with 1 second length in dataset 2 for experiment. Specifically, for SSVEPformer, after the model is trained under the inter-subject rule, it will continue to train for 30 epochs using part of the data from the test subject, and finally validate on the remaining data from that test subject. For FB-SSVEPformer, after the model is trained under the inter-subject rule, it will first use part of the data from the test subject to train each subnetwork for 20 epochs, then train the entire model for 10 epochs, and finally verify it on the remaining data from that test subject. Furthermore, we checked the influence of the different numbers of data blocks at each frequency from the test subject on the classification result for subject-adpative training. The result is shown in the Fig.~\ref{fig:figure6}. It can be seen that using the data from test subject for subject-adaptive can significantly improve the performance of the model, and the accuracy of both models has increased above 90$\%$. Surprisingly, after subject-adaptive with only one block data at each frequency, the performance was improved by 11.80$\%$ and 10.33$\%$ for SSVEPformer and FB-SSVEPformer, respectively. Besides, when the number of the data blocks increases from one to five, the accuracies incease by 2.72$\%$ and 2.36$\%$ for the two methods, respectively. These results show that the proposed models can achieve satisfactory performance when only a small amount of data are available for subject-adaptive classification.
	
	\subsection{Model visualization and interpretation}
	
	\begin{figure*}
		\centering
		\includegraphics[width=4.7in]{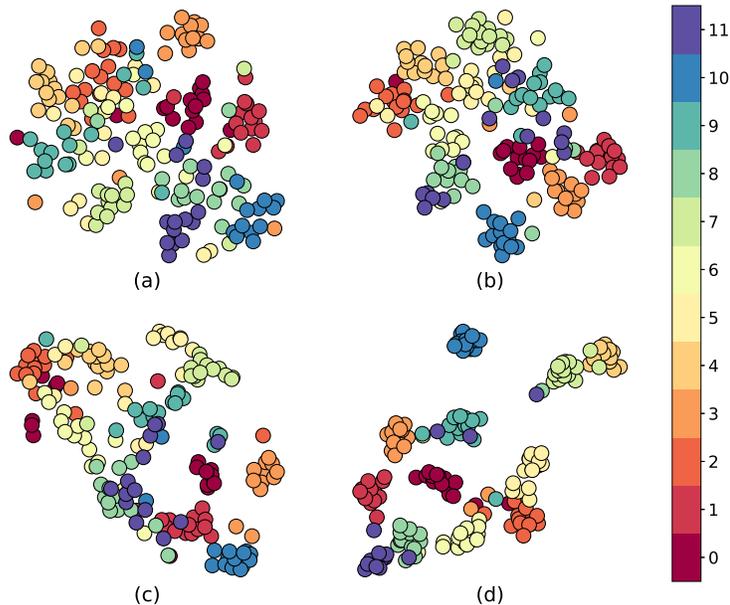} 
		\caption{The t-SNE visualization results of a representative subject with the four models on Dataset 1. (a) EEGNet; (b) CCNN;  (c) SSVEPformer; (d) FB-SSVEPformer. Each dot represents a sample data, and different colors indicate the category to which the data belongs.}
		\label{fig:figure7}
	\end{figure*}
	
	To further display the possible reasons that the proposed methods achieve better results than the baseline methods, we adopted the t-Stochastic Neighborhood Embedding (t-SNE) to  visualize the learned embedding features of the four deep learning methods that yield the top 4 accuracies, i.e, EEGNet, CCNN, SSVEPformer, FB-SSVEPformer. Owing to the large number of categories on Dataset 2 that is hard for plotting the results, we only present the experiment results on Dataset 1 in this section. Fig.~\ref{fig:figure7} shows the visualization results of a representative subject(subject No.3) with the four models on Dataset 1. It can be seen that the features extracted by the proposed SSVEPformer and FB-SSVEPformer have smaller intra-category distance and larger inter-category distance than those of the baseline methods, which can result in better classification results.
	
	\begin{figure*}
	\centering
	\includegraphics[width=4.7in]{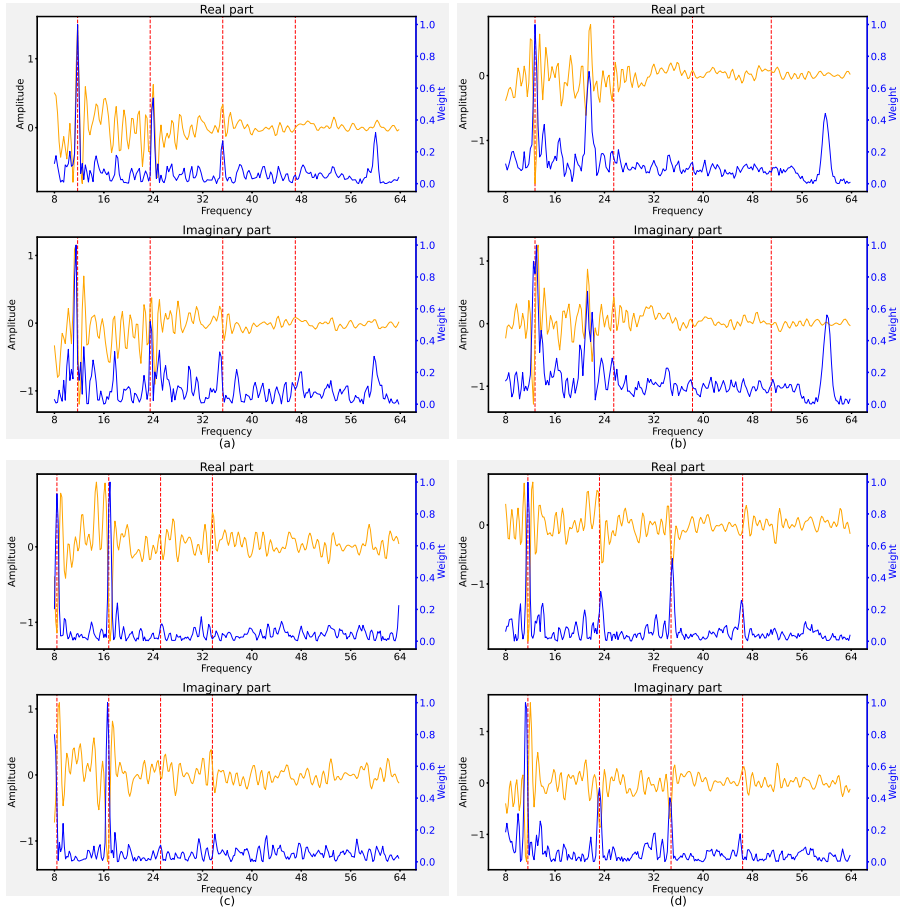}
	\caption{Visualization results of SSVEPformer using grad-CAM, which represent the correlation of each position in the input data with the output of the SSVEPformer encoder. (a) and (b) are grad-CAM visualization results for 11.75 Hz and all 12.75 Hz, respectively, for the a representative subject in Dataset 1. (c) and (d) are grad-CAM visualization results for 8.4 Hz and 11.6 Hz, respectively, for the a representative subject in Dataset 2. The upper and lower subgraphs in (a), (b), (c) and (d) are the results of the real and imaginary parts of the input data, respectively. The yellow line represents the mean of real or imaginary amplitude in the input data of channel Oz. The blue line denotes the weights from the grad-CAM. The vertical red dotted lines indicate the positions fundamental frequency and harmonics.}
	\label{fig:figure8}
	\end{figure*}
	
	The classification mechanism of the deep learning model is still not as intuitive as the traditional machine learning method. So, the model interpretability is an important property for the deep learning method. In current study, we try to adopt the gradient-weighted Class Activation Mapping (grad-CAM) to investigate the interpretability of the SSVEPformer model~\cite{selvaraju2017grad}. Grad-CAM can visually demonstrate how the deep learning model makes the decision based on the input data. The higher the weights, the greater the contribution of the corresponding data of the input data to the result. Concretely, we calculated the heatmaps(weights) with the grad-CAM to represent the relevance of each position in the input data for the output of SSVEPformer encoder. Fig.~\ref{fig:figure8} shows the grad-CAM visualization results of 11.75 Hz and 12.75 Hz for the representative subject (subject No.4) on Dataset 1, and  8.4 Hz and all 11.6 Hz  for the representative subject (subject No.32) on Dataset 2. It can be observed that for both the real part and the imaginary part of the input data, the weights at the stimulation frequency and harmonics points are obviously higher than those of other frequency points. The results of grad-CAM prove that the SSVEPformer can find the classification features in the input data, and use the basic frequency information of the SSVEP data for classification decisions.

	\subsection{Limitation}
	
	Even though both SSVEPformer and FB-SSVEPformer achieve promising performance, some limitations still exist in this study. First, we only used two public datasets to test the model, and  more datasets such as BETA should be adopted in future experiments~\cite{liu2020beta}. Second, although SSVEPformer and FB-SSVEPformer were implemented in inter-subject scenarios, they still require a large amount of data from existing subjects for training. When only limited data are available, how to further compress the amount of training data while maintaining the model performance is a problem worthy of study. Besides, all experiments in this study are performed under offline conditions, and future work can
	further investigate the efficiency and effectiveness in the online SSVEP-based BCI system.
	
	\section{Conclusion}
	
	In the case of time-consuming and laborious data collection, designing a model that can yield excellent classification result under inter-subject conditions is a realistic requirement for SSVEP-based BCI systems. According to the structure of Transformer and the characteristics of SSVEP data, we proposed a SSVEPformer model and its variant FB-SSVEPformer with the filter bank techonolgy. To validate the model performance, we conducted extensive experiments on two public datasets under inter-subject conditions with data lengths ranging from 0.5 s to 1 s. Experimental results show that the proposed model outperforms three popular baseline methods on both datasets. The FB-SSVEPformer model can achieve the best results. Furthermore, we used grad-CAM to visualize the impact of different locations of the input data on the output results, demonstrating the interpretability of the model. The proposed model has high interpretability, and holds promising potential to promote the practical applications of SSVEP-based BCI systems.

	\section*{Acknowledgments}
	This work was supported in part by the National Natural Science Foundation of China under Grant No.62076209 and No.61871423.

   \bibliographystyle{cas-model2-names}	

\end{document}